# Optimization of Reliability of Network of Given Connectivity using Genetic Algorithm

Ho Tat Lam and Kwok Yip Szeto
Department of Physics
The Hong Kong University of Science and Technology,
Clear Water Bay, Hong Kong
phszeto@ust.hk

## ABSTRACT

Reliability is one of the important measures of how well the system meets its design objective, and mathematically is the probability that a system will perform satisfactorily for at least a given period of time. When the system is described by a connected network of $N$ components (nodes) and their $L$ connection (links), the reliability of the system becomes a difficult network design problem which solutions are of great practical interest in science and engineering. This paper discusses the numerical method of finding the most reliable network for a given $N$ and $L$ using genetic algorithm. For a given topology of the network, the reliability is numerically computed using adjacency matrix. For a search in the space of all possible topologies of the connected network with $N$ nodes and $L$ links, genetic operators such as mutation and crossover are applied to the adjacency matrix through a string representation. In the context of graphs, the mutation of strings in genetic algorithm corresponds to the rewiring of graphs, while crossover corresponds to the interchange of the sub-graphs. For small networks where the most reliable network can be found by exhaustive search, genetic algorithm is very efficient. For larger networks, our results not only demonstrate the efficiency of our algorithm, but also suggest that the most reliable network will have high symmetry.

PACS number(s): 89.75.−k, 89.20.−a

## 1. INTRODUCTION

For systems to meet its design objective, reliability of the network describing the various components and their connections is an important measure, which gives the probability that the system will perform satisfactorily for at least a given period of time. Of course, the reliability of the system depends on both the reliabilities of its components and the ways the components are connected. Therefore, one can increase the system reliability by several means. The simplest way is to increase the reliability of the connections between components using parallel redundancy for the links. However, this method to improve system reliability will consume more resource and a better way is needed to achieve a balance between system reliability and resource consumption. If we assume the identical reliability of the connections, we can increase reliability of the network by a suitable choice of connection of the components. This second approach aims at finding the most reliable network among all possible topologies of the network, i.e., all possible ways that all the N components are connected, for a given number L of links without redundancy. This is a mathematically difficult network design problem which solution provides a universal reference for network construction that is relevant in many real world applications, such as telecommunications [1-3], computer networking [4-6], sewage systems, oil and gas lines [7]. Indeed, the importance of network reliability in science and engineering can be seen from the history of journal specializing in reliability, such as IEEE Transaction on Reliability, which has been around for many decades. In the context of statistical physics, one can relate our network design problem is to the percolation processes [8] in complex networks. If one randomly designates each edge either "open" or "closed", then the studies of the various properties of the resulting patterns [9-11] is a topic in percolation. With this in mind, our network design problem is related to the application of bond percolation to the question of network reliability.

Various approaches have been made in the network design problem [12]. One of the most recent one is to consider the reliability problem of network as a problem of network resistance [13], after the discovery of the negative correlation between reliability and network resistance: a smaller network resistance corresponds to a higher reliability [14-16]. One can calculate the importance of a link in terms of the change in the network resistance after its removal, so as to find ways to improve the reliability of the network by protecting those links which removal maximally increases the network resistance. By the combined process of the removal of a link that minimally increases the network resistance and the addition of a link that decreases maximally the network resistance, we can increase the network reliability. Another approach is to consider reliability of the network in the context of epidemiology [17]. Among all these works, we find that our optimization using genetic algorithm can be a practical solution to this very complex mathematical problem of network reliability.

We show in section 2 the definition of the "All Terminal" problem of reliability and the constraint optimization that we will solve numerically using genetic algorithm. In section 3 we describe the method of computing the reliability for a given network. In order to have a benchmark of comparison, we tabulate the reliability of the most reliable network found by exhaustive search in section 4 for small $N$. Section 5 is the main section describing the method of genetic algorithm in search of the topology with highest reliability. We then compare the performance of genetic algorithm with exhaustive search and the time complexity of our algorithm in section 6. More results of larger networks are shown in section 7 with hints on the importance of the symmetry of the graphs in the reliability of the network.

## 2. PROBLEM DESCRIPTION

We focus on the design of network assuming that the system consists of $N$ components and there are exactly $L$ links provided for connecting them. Between any two components, there can either be one or zero link connecting them. Here we assume every link of the network has same reliability, measured by a parameter $p$ which is the probability that the link is working (and probability $q=1-p$ that the link is broken). This $p$ is called the robustness parameter. This robustness parameter is the same as the bond open probability $p$ in bond percolation, where we assume that every link of the network are randomly designated either "open" (successful) or "closed" (failed), with probability $p$ that a link is designated "open".

We follow the well-established framework in the theory of signature [18] to analyze network reliability. The success of the system in terms of reliability means that every pair of components (nodes in the network language) has at least one path connecting them (called the path-connectedness). Reliability of the system is thus a probability $R$ of success of the working of the system for a given robustness parameter $p$. In layman terms, network reliability is the probability that the network is connected: every node can communicate with any other node. This description of the reliability of the network is called the "All Terminal" reliability as it requires the network consists of one single connected component. There are other measures of reliability such as two-terminal reliability, or the "attack rate" reliability used in epidemiology [17]. In the context of "All Terminal" reliability, the optimization problem of the reliability of a network is equivalent to the following design problem: given $N$ nodes and $L$ links, find the most reliable topology. In this paper, we assume our network is the simplest undirected un-weighted network. If our network is fully connected, then there are $L_{\max} = N(N-1)/2$ links. On the other hand, a connected network with $N$ nodes requires at least $N$-1 links, which gives a tree structure for the topology. Thus, for a general network of $N$ nodes, the number of links $L$ is between $N$-1 and $N(N-1)/2$. The connectivity of the network is defined by $\rho \equiv L/L_{\max} = 2L/(N(N-1))$. The network design problem is to find among $C_L^{N(N-1)/2}$ possible graphs the most reliable one [12,18,19] for given $N$ and $L$. One sees that the solution space is very large and our problem of network reliability is a very difficult combinatorial optimization problem [20]. Moreover, the evaluation of reliability of a given network, when it is not the usual series or parallel configuration, is difficult as it depends on the robustness parameter as well as the topology of the network. One can show that the reliability of a network can be expressed as a polynomial in the robustness parameter after the famous work by Moore and Shannon [21]. For a given topology of network, the probability of the system working is a monotonic function of the robustness parameter, with a general shape like the Fermi function. Reliability of the network is zero when $p=0$ and it is 1 when $p=1$. In this paper, we assume that the robustness parameter $p$ is set to 0.7 in our numerical simulation of the reliability of a given network. Our focus is on the topology of the network.

In general, we expect that exhaustive search for the most reliable network for large $N$ is impossible computationally. Therefore, we set our objective on the algorithmic aspect of finding the most reliable network numerically. Hopefully from our numerical work we can gain some insights into the general feature of the topology of the most reliable network. Our aim is to design a genetic algorithm to search for the most reliable network of given $N$ and $L$, for large $N$.

## 3. RELIABILITY EVALUATION

Before we can search for the best topology in terms of reliability, we need to evaluate the network reliability $R$ for a given robustness parameter $p$ and a given topology. As we focus on the simplest undirected and un-weighted networks with no redundant links between two nodes, we can describe their topology with a special adjacency matrix **A**. Initially, all the nodes of the network are labeled from 1 to $N$ and the adjacency matrix **A** of a network with $N$ nodes is a $N \times N$ matrix with entry $A_{ij}=1$ if the node $i$ and node $j$ are connected by a link, and $A_{ij}=0$ otherwise. This matrix representation is useful in evaluating the network reliability numerically when we incorporate the robustness parameter into the network. In order to evaluate the reliability of the network with adjacency matrix **A** and robustness parameter $p$, we introduce a random matrix $\mathbf{A}_{rand}$ in Algorithm 1 to compute the reliability $R=R(p;N,L)$ of **A** using Monte Carlo Method.

---
**Algorithm 1** Evaluate network reliability for a given matrix **A** with a given robustness
---
n_exp: = number of experiment     n_success:= 0
**for** n = from 1 to n_exp **do**
let $\mathbf{A}_{rand}$ be a random N×N matrix whose entries are random number in the interval [0,1]
**if** the $ij^{th}$ entries of $\mathbf{A}_{rand}$ is smaller than $p$ and the corresponding $ij^{th}$ of **A** is 1
  **then** let the $ij^{th}$ entries of $\mathbf{A}_{rand}$ be 1
  **else** let the $ij^{th}$ entries of $\mathbf{A}_{rand}$ be 0
**if** $\mathbf{A}_{rand}$ is connected
  **then** let n_success = n_success + 1
**end for**
$R$ = n_success/n_exp

---

In this algorithm, we need to check if the network with adjacency matrix **A** is connected, in other word, its path-connectedness. The $ij^{th}$ entry of $\mathbf{A}^n$ is the number of paths of length $n$ connecting node $i$ and node $j$. Thus, when the $ij^{th}$ entry of $\mathbf{A}^n$ is 0, there is no path of length $n$ connecting node $i$ and node $j$ and we say that node $i$ and node $j$ are not path connected by a path of length $n$. If node $i$ and node $j$ are not path connected of length $n$ for all possible $n$ (which can be 1,2,… N, since the maximum length of path between two nodes is N), then we can say that node $i$ and node $j$ has no connected path (i.e., node $i$ and node $j$ are *not* path-connected). Mathematically, this means that when the $ij^{th}$ entry of the summation of $A^i$ for $i$ running from 1 to $N$ is zero, node $i$ and node $j$ are *not* path connected. The number of paths between node $i$ and node $j$ is given by the $ij^{th}$ entry of the following matrix

$$A_{path} = \sum_{i=1}^{N} A^i = (I - A^{N+1})(I - A)^{-1}$$

where $I$ is an N×N identity matrix. To check if a network is connected, we check for its path-connectedness by Algorithm 2.

---
**Algorithm 2** Check the path-connectedness of the network
---
**if** (I-A) is not singular
  **then** let $\mathbf{A}_{path}$ be the matrix $(I - A^{N+1})(I - A)^{-1}$
  **else**     $\mathbf{A}_{path}$ := summation of $A^i$ for $i$ from 1 to $N$
**if** all entries of $\mathbf{A}_{path}$ are nonzero
**then** the network is connected

---

After using Algorithm 2 to check if **A** in Algorithm 1 is connected, we can complete all steps in Algorithm 1 and obtain the reliability of the given network **A**.

Note that the evaluation of network reliability contains random error. This error can be reduced by increasing the number of experiment, n_exp. From these two algorithms, we see that the evaluation of reliability consumes substantial computational resource.

## 4. EXHAUSTIVE SEARCH

A key step of our genetic algorithm is to generate a new network, which forms the "*chromosome*" in our evolutionary computation, and evaluating its reliability, which is the fitness function evaluation. In order to assess the performance of genetic algorithm in finding the most reliable network, we need to make comparison with the known results from exhaustive search.

---
**Algorithm 3** Exhaustive search
---
G: = {All networks with N nodes and L links}
**for** all networks in G **do**
    evaluate network reliability of the network
**end for**
rank the networks and find the most reliable network

---

The results of the exhaustive search for $N$ = 4, 5 and 6 and different $L$ are shown in Table 1. The robustness is set to be 0.7 and n_exp is 1000.

**Table 1.** The network reliability of the most reliable network with given N and L when the robustness is set to be 0.7. (Computed using exhaustive search)

| N | L | Reliability | N | L | Reliability |
|---|---|---|---|---|---|
| 4 | 4 | 0.639 ±0.032 | 4 | 5 | 0.792 ±0.032 |
| 4 | 6 | 0.899 ±0.032 | 5 | 5 | 0.528 ±0.032 |
| 5 | 6 | 0.703 ±0.032 | 5 | 7 | 0.798 ±0.032 |
| 5 | 8 | 0.893 ±0.032 | 5 | 9 | 0.941 ±0.032 |
| 5 | 10 | 0.960 ±0.032 | 6 | 6 | 0.4193±0.032 |
| 6 | 7 | 0.621±0.032 | 6 | 8 | 0.733±0.032 |
| 6 | 9 | 0.853±0.032 | 6 | 10 | 0.888±0.032 |
| 6 | 11 | 0.926±0.032 | 6 | 12 | 0.959 ±0.032 |
| 6 | 13 | 0.966 ±0.032 | 6 | 14 | 0.975 ±0.032 |
| 6 | 15 | 0.985 ±0.032 | 7 | 10 | 0.720 ± 0.001 |

However, exhaustive search requires searching over all possible networks. As we noted in the introduction, the space of all possible networks with $N$ nodes and $L$ link has $C_L^{N(N-1)/2}$ members, or steps, which is of the order of $O\left(\left[(1-\rho)^{1-\rho}\rho^\rho\right]^{N(N-1)/2}\right)$ where the connectivity of the network is $\rho \equiv L/L_{max} = 2L/(N(N-1))$. We see that this is an astronomical number for large *N* and nontrivial connectivity, making exhaustive search impossible in practice. (Note that the trivial case of zero connectivity is that the network has no link and cannot be reliable. When connectivity is one, the network is always reliable.) Therefore, we limit ourselves for comparison in networks with small *N* first and intermediate connectivity first.

## 5. GENETIC ALGORITHM

Here, a general method of applying genetic algorithm in graph searching is introduced. Our genetic algorithm in searching for the most reliable network is shown in Algorithm 4. The fitness measure is the network reliability. Each candidate network is a chromosome in a population. We select the fittest solution, which is the network with highest reliability, and change the other networks in the population by three different operators: creation, mutation and/or crossover with the fittest chromosome. In our genetic algorithm, the termination condition determines the quality of the output of the algorithm. With no information of the exact solution when *N* is large, we use the following termination condition: when the most reliable network does not change for n_tor generation, the algorithm will terminate. Here we can increase the quality of the output by increasing n_tor.

| **Algorithm 4**   Genetic algorithm |
|---|
| n_pop: = number of networks in a generation |
| generate a population with n_pop networks |
| evaluate reliability of the networks in the generation |
| **while** the condition is not satisfied |
| select the population and produce new generation |
| evaluate reliability of the networks in new generation |
| find the most reliable network in the generation |
| **end while** |
| output the most reliable network of the last generation |

## 5.1 GENETIC REPRESENTATION

While the representation of the network by adjacency matrix is useful for the evaluation of network reliability, it is not as useful in the implementation of genetic algorithm, where the chromosome is a string. Here we first introduce a mapping from the adjacency matrix to a string. Initially, the entries in the upper triangle of the adjacency matrix are labeled row by row, starting from the first to the last row. Since we address only undirected networks, their associated adjacency matrices are symmetrical with only $L_{max} = N(N-1)/2$ independent entries in the N×N adjacency matrix **A**. It means that the upper triangle of **A**, if labeled by row sequentially, corresponds to a binary string of length $L_{max}$. The entries of this long representation are either 0 or 1. For example, network (a) in Fig.1 can be represented by the binary string 1001111101. However, if we focus on network with only *L* links, then we see that among this string of length $L_{max}$, there are *L* positions that are one, and the remaining $L_{max} - L$ positions are zeros. Therefore, we can simplify the string representation of the network using a shorter string of length *L*, so that the network is uniquely represented by a string of *L* numbers recording the position of the nonzero entries in the matrix. As an example, we illustrate the string representations and the adjacency matrices of three networks with 5 nodes and 7 links in Fig.1. We see that the binary string representation in Fig.1(a) by 1001111101 can now be replaced by a length of L=7 units, : (1, 4, 5, 6, 7, 8, 10). Here the number in each entry ranges from 1 to $L_{max}$.

```
          NETWORK       ADJACENCY       STRING
                         MATRIX     REPRESENTATION
               1
         5                0 1 0 0 1
                          1 0 1 1 1
(a)                2      0 1 0 1 0     1 4 5 6 7 8 10
                          0 1 1 0 1
         4                1 1 0 1 0
               3

                          0 1 1 0 1
                          1 0 0 1 0
(b)                       1 0 0 1 1     1 2 4 6 8 9 10
                          0 1 1 0 1
                          1 0 1 1 0

                          0 1 1 0 1
                          1 0 1 1 0
(c)                       1 1 0 1 0     1 2 4 5 6 8 10
                          0 1 1 0 1
                          1 0 0 1 0
```

**Fig.1.** Networks with N=5 nodes and L=7 links with their adjacency matrices and short string representations. The nodes of all networks are labeled in the clockwise direction, as shown in network (a). The number in the short string representation corresponds to the ordering of the string in the long representation of length $L_{max}$ =10. For example, network (a) has binary string representation 1001111101 of length 10 and its short string representation of length 7 is (1 4 5 6 7 8 10). Similarly the binary string representation for (b) is 1101010111 with the associated short string (1 2 4 6 8 9 10) and for (c) is 1101110101 with the short string representation (1 2 4 5 6 8 10)

## 5.2 GENETIC OPERATOR

The new generation of chromosomes (networks) in our population is produced by three methods: creation, mutation and crossover. These three operators have different effects on the search process. Creation creates a new network. It is used to generate the initial population and replace those networks with relatively low reliability in the subsequent evolution. Mutation changes one of the entries of the string. In the network language, it removes one of the links in the network and places it at another position corresponding to rewiring of the network. The new network obtained from mutation has similar reliability as the original one. Thus, mutation operator allows us to perform local search for higher fitness, i.e., rewiring to improve the reliability of network. Crossover finds the two disjoint sets of two strings, divides the set into two pieces and interchanges these pieces of the strings. It can be single or multiple point crossover. In other words, crossover divides the networks into two subgraphs, retains the connections between them but interchanges the subgraphs. Crossover allows those reliable structure interchanges within the population. Let us illustrate these genetic operators with some examples. By changing the fifth entry, 7 of the short string representation of network (a) into 2, the original network a=(1001111101) or (1 4 5 6 7 8 10) is rewired into c=(1101110101) or (1 2 4 5 6 8 10). This corresponds to have a mutation on the fifth entry of the short string representation of (a). On the other hand, (c) can also be obtained by performing crossover of (a) with network (b)= (1101010111) or (1 2 4 6 8 9 10). We observe from their binary string representation that the disjoint set of (a) and (b) consists of 2,5,7,9. Selecting the crossover points at 2 and 7, two offspring networks, (c) and another network (1001011111) or (1 4 6 7 8 9 10) are obtained. This example shows that we have at least two ways to obtain (c) with different genetic operators on (a) or (b).

## 5.3 SELECTION AND PRODUCTION

We first perform a set of experiments to compare the efficiency of the three genetic operators: creation, mutation and crossover. We address the problem of finding the most reliable network(s) with 7 nodes and 10 links whose solution space contains 352716 networks. From exhaustive search, we find three solutions shown in Fig.2. They all have similar reliability of 0.720 when the robustness parameter $p$ is 0.7. Using genetic algorithm to find networks of high reliability, we starts with a population of five chromosomes which are five networks with 7 nodes and 10 links. In each generation, we eliminate the least reliable network and let the remaining population of networks be the "old generation", which now contains only four networks. To form the new generation of networks, we keep the most reliable network from the old generation and then apply different *"appropriate set of genetic operators"* to generate another four new chromosomes. In Table 2, we list the models, together with the associated *"appropriate set of genetic operators"*, and the resultant reliability of the best network after 10 generations of evolution.

In order to compare their performance, we allow each model to run for 10 generations using the specified genetic operators on the population before registering the fittest chromosome (the most reliable network). Among the five models listed in Table 2, the first four models are used to compare the efficiency of the three operators and the fifth model is our chosen model for finding the most reliable network with larger size. Our numerical results for the experiment on small network suggest that mutation is the most efficient genetic operator and crossover reduced the efficiency of the algorithm whereas creation is least efficient for this chosen problem with small $N$ and $L$.

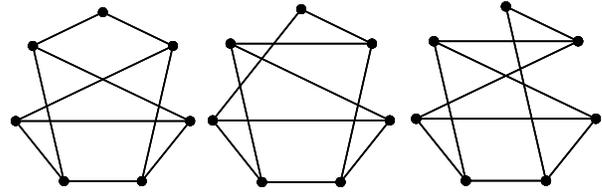

**FIG. 2.** The most reliable network with 7 nodes and 10 links. Their reliability is 0.720 when the robustness is set to be 0.7.

**Table 2.** The reliability of the best network after 10 generations. There are five models defined by their different methods of production of the other four chromosomes in the new generation of network. The percentage represents the probability of using the corresponding operators.

| Models | Method of production of four networks for the new generation | Network reliability |
|---|---|---|
| 1 | Creation | 0.6874 |
| 2 | Mutation | 0.7039 |
| 3 | 90% mutation + 10% crossover | 0.6913 |
| 4 | 50% mutation + 50%crossover | 0.6698 |
| 5 | 90% mutation + 10% crossover + replace the worst network by creation | 0.7220 |

We now explain the experimental results heuristically. We see that even for our small network with N=7 and L=10, the solution space of the problem and the available position of links are already quite large. However, our population of chromosomes is small, consisting of only five networks. Mutation is more efficient in producing better networks because mutation used the information of the most reliable network in the old population whereas creation does not. Nevertheless, creation provides diversity of search area, as it generates new network randomly. This prevents the search process from getting trapped in a local region of the solution space. Similarly, crossover may displace the search to another sector of the solution space, though it can avoid the search process from getting trapped in a local region. It differs from creation in that it still keeps some of the features of existing chromosomes (some subgraphs).

Based on the characteristics of the three operators, we argue in favor of the following strategy.

(1) Reserve the most reliable network in the population, (keep the fittest chromosome).

(2) Use mutation and crossover with the most reliable network on the second and third networks to generate two new chromosomes.

(3) The fourth chromosome is obtained by rewiring the most reliable network.

(4) The fifth chromosome is replaced by creation. In the rest of this paper, we only apply this method to produce the new generation.

For generalization to a larger population of chromosomes, we may incorporate more combination of these genetic operators, or introducing some new genetic operators. This may further increase the efficiency of our algorithms. For the sake of simplicity, the above formulation using only five chromosomes suffices to demonstrate the power of genetic algorithm in solving this rather difficult optimization problem in networks.

## 6. PERFORMANCE AND ANALYSIS

To verify the efficiency of genetic algorithm, we conduct a set of experiment to search for the most reliable network for given N and L.

### 6.1 NUMERICAL COMPARISON

First we perform exhaustive search to obtain the most reliable network(s). Next we run our genetic algorithm with preset termination condition. For example, we stop the search once the fittest chromosome matches the most reliable network found from exhaustive search. The number of generations used is recorded. We then repeat this process ten times using genetic algorithm, obtaining ten different number of generation needed, from which we record the maximum number. Due to the computation effort required in the evaluation of reliability of a given network, we only limit ourselves to ten runs of our genetic algorithm. Ideally we may repeat many more times, such as one thousand runs, in order to obtain enough statistics for the first passage time distribution of the number of generation required to first match the most reliable network found from exhaustive search. However, with our limited runs we can already demonstrate the efficiency of our algorithm. We will provide a better analysis of the time complexity later to further validate our claim without using the first passage time analysis.

In order to compare the searching time of genetic algorithm with exhaustive search, we use the measurement unit called the "*Computational unit*". Here each computational unit stands for the computational effort required in generating a new network and evaluating its reliability. In our genetic algorithm, we assume that generating networks by creation, mutation or crossover takes approximately the same computational effort, so that the searching time is not the actual CPU time but the number of networks that the algorithm required to try before reaching the exact solution, i.e., the most reliable network for the given robustness parameter *p*. Note that the major computational cost is incurred in the evaluation of the reliability, not the generation of the network for test. From the results summarized in Table 3, our experiments on small network (N=6,L=7) show evidence that genetic algorithm is more efficient than exhaustive search.

**Table 3.** The first generation when the algorithm finds the exact solution for given N and L. The value in the ratio column is the ratio of the computational unit for search using genetic algorithm and exhaustive search. The number in the parentheses is the population size.

| N | L | Generation | Ratio | N | L | Generation | Ratio |
|---|---|---|---|---|---|---|---|
| 6 | 7 | 31(5) | 0.019 | 6 | 8 | 72(5) | 0.045 |
| 6 | 9 | 277(5) | 0.222 | 6 | 10 | 166(5) | 0.221 |
| 6 | 11 | 17(5) | 0.051 | 7 | 10 | 468(10) | 0.012 |

We introduce a ratio for comparison of our genetic algorithm with exhaustive search in Table 3. This is the ratio between the numbers of the networks that two algorithms required to reach to exact solution. For genetic algorithm with a population of five chromosomes, each generation need to generate four networks in our setup of the population. Thus, in Table 3, the case of N =6, L = 7 requires 31 generation, implying that our genetic algorithm needs to generate 5 chromosomes for the initial population, and afterwards 4 per new generation (30 of them). The total number of chromosomes (networks) computed is thus 5+4×30, while the exhaustive search in general needs approximately $C_7^{15}$ = 6435 networks. Note that in exhaustive search, we need to use all $C_7^{15}$ networks before we can say that we have found the most reliable network. In the case of genetic algorithm, we can never be sure that the fittest chromosome is the most reliable network, as we can only say that it is the fittest up to the number of generations used. However, if we look at the quality of the solution found from genetic algorithm, and compare it with the exact solution of the most reliable network found from exhaustive search, we can say that our comparison on the efficiency of genetic algorithm with exhaustive search is a fair comparison and the ratio listed in Table 3 is a good indicator that genetic algorithm is quite efficient in solving this reliability problem of networks.

### 6.2 TIME COMPLEXITY ANALYSIS

Now we analyze the time complexity of our algorithm in the search for the most reliable network with *N* nodes and *L* links. We can first map the solution space of this problem into a graph which vertices represent the chromosomes (networks). We define the edge connecting two vertices by the condition that the two chromosomes can be transformed into each other by the rewiring of one link. There are $C_L^M$ networks with *N* nodes and *L* links, where $M = L_{max} = N(N-1)/2$. Among these *M* entries, there are *L* of them assuming the values of 1 and the remaining being 0. Therefore, there are $C_L^M$ possible combination, putting *L* 1s into *M* entries. This then implies that there are $C_L^M$ vertices in our graph for the solution space. Starting from a network of *N* nodes and *L* links, it can reach other network with the same *N* and *L* by rewiring. The number of possible rewiring is determined by the number of empty spot (the number of zero) in the adjacency matrix, which is *E=M-L*. As there are *L* links to be chosen for rewiring, the total number of possible networks reachable by rewiring is simply *L×E* starting from a given network. This shows that our graph representation of the solution space is a regular graph of $C_L^M$ vertices and each vertex has degree *L×E*. Since networks with similar structure have similar reliability, two adjacent vertices in the graph of the solution space correspond to two networks with similar reliability. This graphical representation of the solution space indicates that the problem of finding the network with maximum reliability is actually soluble by a hill climbing method, as the solution space is "smooth" in sense of the function that we called reliability, even though this graph is not the usual hypercubic lattice. To obtain a satisfactory solution, the algorithm only needs to search in the local neighborhood of the initial network and move towards the adjacent vertices that have a higher reliability. This then explains the advantage of the model that uses 90% mutation in Table 2, since mutation of the string representation is a simple rewiring of the network. Searching in the neighborhood of a given vertex requires reaching all nearest neighbors by rewiring, which means *L×E* choices are made. Thus, the order of magnitude of operations required to find the most reliable network in the neighborhood of a given vertex is O(*EL*). Since the graph has a radius *L* (any two vertices can be reached by at most L steps of rewiring), we expect that *L* steps is required to reach the furthest vertex in graph of the solution space, the number of steps to find the most reliable network with *N* nodes and *L* links is therefore of the order
$O(L(LE))=O(L^2(M-L)) = O(L^2(N(N-1)/2 - L))$
which is a polynomial of *L* and *N*.

## 7. SEARCH FOR LARGE NETWORK

In the previous sections, we develop an efficient method in the search for the most reliable network of fixed connectivity and small N. In this section, we apply the algorithm to larger networks. We focus on the networks with 10 nodes where computation is not yet too expensive but the result is representative. The solution space of networks with 10 nodes is of order of $C_L^{45}$. We first study the networks whose number of links *L* is the multiple of the half of the number of nodes *N,* implying that the solution space corresponds to a graph with $C_{L=15}^{M=45} \approx 3.4 \times 10^{11}$ vertices for *L=15*. The most reliable networks along with their adjacency matrices obtained from our genetic algorithm are shown in Figure 3-8.

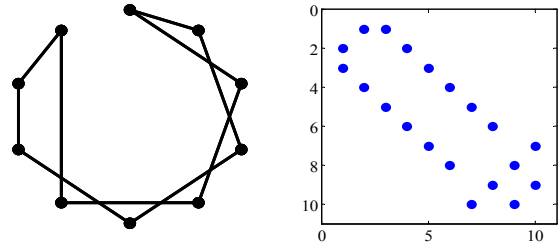

FIG.3. The searching result for 10 nodes and 10 links. Its reliability is 0.748 when the robustness is set to be 0.9.

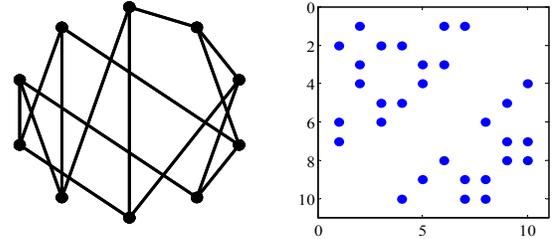

FIG.4. The searching result for 10 nodes and 15 links. Its reliability is 0.734 when the robustness is set to be 0.7.

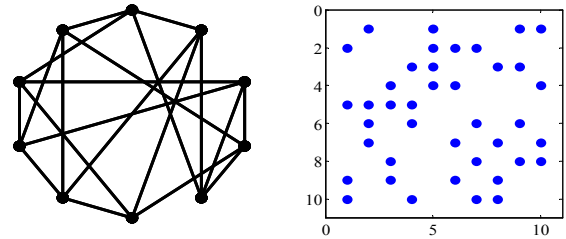

Figure5. The searching result for 10 nodes and 20 links. Its reliability is 0.932 when the robustness is set to be 0.7.

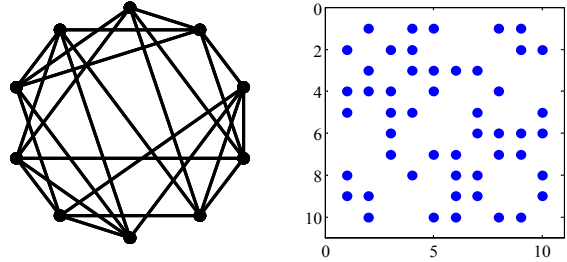

FIG.6. The searching result for 10 nodes and 25 links. Its reliability is 0.737 when the robustness is set to be 0.5.

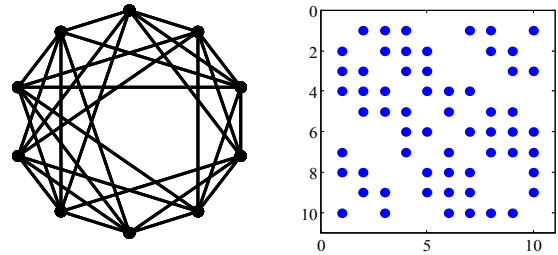

FIG.7. The searching result for 10 nodes and 30 links. Its reliability is 0.882 when the robustness is set to be 0.5.

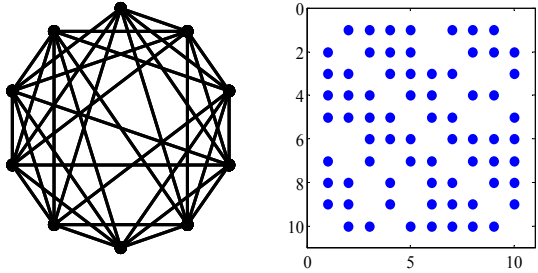

FIG.8. The searching result for 10 nodes and 35 links. Its reliability is 0.745 when the robustness is set to be 0.4.

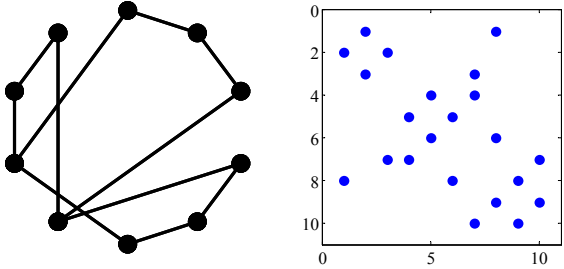

FIG.9. The searching result for 10 nodes and 11 links.

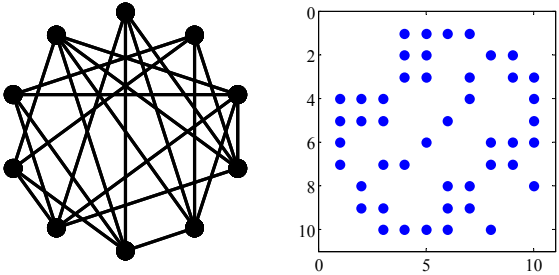

FIG.10. The searching result for 10 nodes and 23 links.

The reason that we choose *L* being multiple of *N/2* is that they can form regular networks, which nodes have same degree, i.e., same number of neighboring nodes. According to the search results from our genetic algorithm, the most reliable networks with 10 nodes and links of the multiple of 5 are all regular networks. Furthermore, regular networks obtain the highest symmetry in terms of degree distribution among all networks with same connectivity.

Since the reliability is a function of the robustness parameter *p* that behaving like a Fermi function, with reliability being 0 at *p*=0 and 1 for *p*=1. There exists in general a region of *p* where the reliability increases rapidly from zero to one. It is in this region of *p* that our detection of the effect of topology of the network on reliability can achieve highest sensitivity. Therefore in Fig.3-8, we perform our comparison of reliability for different topology at a chosen robustness value, where reliability is most sensitive to topology. Note that Fig.3-8 shows the most reliable network topology for a given N and L. In making the claim that they are the most reliable, it means that we compare many different topologies of networks with the same N and L and also with a fixed robustness *p*. Since the value of *p* used to achieve the most sensitive comparison of the reliability of networks is different for different set of (N,L), we have to adjust the robustness parameter *p* accordingly.

We also search for those pairs of *N* and *L* which do not form regular network. If we look at the degree distribution of the most reliable network found by genetic algorithm, we see that in Fig.9 (*L*=11) and Fig.10 (*L*=23), that they are all very narrow, though it is not the delta function for the degree distribution of the regular network.

Based on these numerical results, we make the following conjecture: for all pairs of *N* and *L*, the most reliable network has the most concentrated degree distribution. In other words, the nodes of the most reliable network tend to have the same degree. We further conjecture the network with higher symmetry should have higher reliability.

## 8. CONCLUSION AND DISCUSSION

In conclusion, we have developed a general method to find the most reliable network with N nodes and L links using genetic algorithm. Within the confine of our numerical experiments, we show that the efficiency of genetic algorithm is much higher than exhaustive search, and is polynomial in N and L in time complexity. We also argue that the most reliable network of N nodes and L links has the most concentrated degree distribution, closest to a regular network which is most symmetrical.

Although our algorithm provides an efficient way to search the most reliable networks, it still requires huge computational efforts due to the expensive evaluation of reliability. Our results can only be treated as approximate due to the intrinsic error in the evaluation on reliability. However, when the network size is large, the difference in the reliability of two networks with same N and L is small and the search for the most reliable network inevitably requires a larger number of experiments (n_exp) used in Algorithm 1 to ensure the accuracy of the comparison. The evaluation of network reliability could be improved by accurately calculating it through considering all of the possible $2^L$ cuts on the network and checking their connectivity. However, based on our guideline on the symmetry of the most reliable network, we can first construct an ansatz using a regular network plus any extra links that are needed to satisfy constraint on the total number of links=L, followed by a search using rewiring, or using our genetic algorithm.

We should mention here that analytical work for reliability is still lacking for large network, even though our numerical method using genetic algorithm provides solution with good quality with limited computational resource. It is our aim to provide theoretical guideline in the construction of the most reliable network. We expect that symmetry analysis will facilitate this research, as our numerical results for small network of size N up to 10 indicate that the most reliable network does have high symmetry, with narrow degree distribution. A combination of the genetic algorithm and the method of comparison of the reliability of two large networks using eigenvalue spectrum of the adjacency matrix [22] will be very powerful in solving the network design problem.

Finally, we note that our algorithm should be complementary to some existing methods of search for reliable networks, such as those using simulated annealing [1,3,5], tabu search [23,24], and other genetic algorithms [12,25-28], as our method can be readily extended to larger networks. In terms of application, our method should be useful also in other problems on network searching when we change the fitness function, such as in the problems on traffic, circuit design and communication network [29].

## 9. ACKNOWLEDGEMENT

K.Y. Szeto acknowledges the support of grant FS-GRF13SC25 and FS-GRF14SC28. We acknowledge many useful discussions with Degang Wu and Zitao Wang.

## 10. REFERENCE

[1] Aiiqullah, M.M. and Rao, S.S. (1993) Reliability optimization of communication networks using simulated annealing. Microelectronics and Reliability, 33, 1303-1319.
[2] Jan, R.-H., Hwang, F.-J. and Chen, S.-T. (1993) Topological optimization of a communication network subject to a reliability constraint. IEEE Transactions on Reliability, 42, 63-70.
[3] Pierre, S., Hyppolite, M.-A., Bourjolly, J.-M. and Dioume,O. (1995) Topological design of computer communication networks using simulated annealing. Engineering Applications of Artificial Intelligence, 8, 61-69.
[4] Aggarwal, K.K., Chopra, Y.C. and Bajwa, J.S. (1982) Topological layout of links for optimizing the overall reliability in a computer communication system. Microelectronics and Reliability,22, 347-351.
[5] Fetterolf, P.C. and Anandalingam,G. (1992) Optimal design of LAN-WAN internet works: an approach using simulated annealing. Annals of Operations Research, 36, 275-298.
[6] Wilkov, R.S. (1972) Design of computer networks based on a new reliability measure, in Proceedings of the Symposium on Computer-Communications Networks and Teletraffic, Fox, I.(ed.), Polytechnic Institute of Brooklyn, Brooklyn, NY. pp. 371-384.
[7] Walters, G.A. and Smith, D.K. (1995) Evolutionary design algorithm for optimal layout of tree networks. Engineering Optimization,24, 261-281
[8] D. Stauffer, A. Aharony, Introduction to Percolation Theory (Taylor & Francis, London, 1994)
[9] M.E.J. Newman, SIAM Rev. 45, 167 (2003)
[10] R. Cohen, S. Halvin, Complex Networks: Structure, Robustness and Function (Cambridge University Press, Cambridge, 2010)
[11] M.E.J. Newman, Networks: An Introduction (Oxford University Press, Oxford, 2010)
[12] Kuo, W., Prasad, V. R., Tillman, F. A., Hwang, C.: Optimal Reliability Design, Cambridge University Press, Cambridge, MA (2001)
[13] Xiangrong Wang, Evangelos Pournaras, Robert E. Kooij, and Piet Van Mieghem, Eur. Phys. J. B (2014) 87: 221
[14] W. Ellens, F.M. Spieksma, P. VanMieghem, A.Jamakovic, R.E. Kooij, Linear Algebra Appl. 435, 2491 (2011)
[15] A. Asztalos, S. Sreenivasan, B.K. Szymanski, G. Korniss, Eur. Phys. J. B 85, 288 (2012)
[16] W. Abbas, M. Egerstedt, Robust Graph Topologies for Networked Systems, in 3rd IFAC Workshop on Distributed Estimation and Control in Networked Systems, 2012, pp. 85-90
[17] Mina Youssef, Yasamin Khorramzadeh, and Stephen Eubank Phys. Rev. E 88, 052810, 2013
[18] Francisco J. Samaniego,System signatures and their applications in engineering reliability, Boston, MA: Springer Science+Business Media, LLC, 2007
[19] Colbourn, C. J.: The Combinatorics of Network Reliability. Oxford University Press, Oxford (1987)
[20] Garey, M.R. and Johnson, D.S. (1979) Computers and Intractability: A Guide to the Theory of NP-Completeness, W.H. Freemanand Co.. San Francisco, CA.
[21] E. Moore and C. Shannon, J. Franklin Inst. 262, 191 (1956).
[22] Zitao Wang and Kwok Yip Szeto Eur. Phys. J. B (2014) 87: 234
[23] Glover F., Lee, M. and Ryan, J. (1991) Least-cost network topology design for a new service: an application of a tabu search. Annals of Operations Research, 33, 351-362.
[24] Koh, S.J. and Lee, C.Y. (1995) A tabu search for the survivable fiber optic communication network design. Computers and Industrial Engineering, 28, 689-700.
[25] Dengiz, B., Altiparmak F. and SmithA.E. (1997) Efficient optimization of all-terminal reliable networks using an evolutionary approach. IEEE Transactions on Reliability, 46, 18-26.
[26] Kumar, A., Pathak R.M., Gupta, Y.P. and Parsaei, H.R. (1995) A genetic algorithm for distributed system topology design. Computers and Industrial Engineering, 28. 659670.
[27] Kumar, A., Pathak, R.M. and Gupta, Y.P. (1995) Genetic-algorithm-based reliability optimization for computer network expansion. IEEE Transactions on Reliability, 44, 63-72.
[28] Deeter, D.L. and Smith, A.E. (1997) Heuristic optimization of network design considering all terminal reliability, in Proceedings of the Reliability and Maintainability Symposium. IEEE, Piscataway, NJ. pp. 194-199.
[29] Aggamal,K.K. and Rai,S.(1981)Reliability evaluation in computercommunication networks. IEEE Transactions on Reliability,R-30, 32-35.